\documentstyle[11pt,newpasp,twoside,epsf]{article}
\begin{document}
\def\vec#1{\mathbf{#1}}

\title{Orbits from Hipparcos}
\author{D.~Pourbaix}
\affil{Research Associate at FNRS, Institute of Astronomy and Astrophysics, Universit\'e Libre de Bruxelles CP 226, Bld du Triomphe, B-1050 Bruxelles}
\begin{abstract}
Among the 120\,000 objects in the Hipparcos catalogue, 235 were fitted with
an orbital model, i.e. with up to seven additional parameters with
respect to the default single star model.  In their quest for the orbital
inclination, spectroscopists promptly realized how useful the Hipparcos
data could be.  Besides the original 235 systems, most Hipparcos entries
with a spectroscopic orbit (extrasolar planet or stellar companion) have 
now been re-processed.  Not all these revised fits were fruitful.  Some 
were even awful.  We present a survey of all the areas where the 
Hipparcos observations have been fitted with an orbital model so far.

\end{abstract}

\section{Introduction}
Besides the five basic astrometric parameters (position, parallax and proper motion), the Hipparcos Intermediate Astrometric Data (IAD) can also yield the parameters of an astrometric orbit when the same IAD are fitted accordingly.  Whereas the binarity is a necessary condition for applying the orbital model, it is far from being a sufficient one.

After a brief memento about the Hipparcos data and the way to model them, we present cases where the orbital model was essentially used to improve the parallax or the proper motion.  In Sect.~\ref{Sect:Grail}, we then focus on the derivation of the inclination.  We will show that the binarity, whether the companion is stellar or substellar, is not a sufficient condition to apply the orbital model.  We will also briefly mention (Sect.~\ref{Sect:Searching}) some recent studies where the Hipparcos data are used to identify new binaries. 

\section{Memento}\label{Sect:Memento}
\begin{figure}[htb]
\plotfiddle{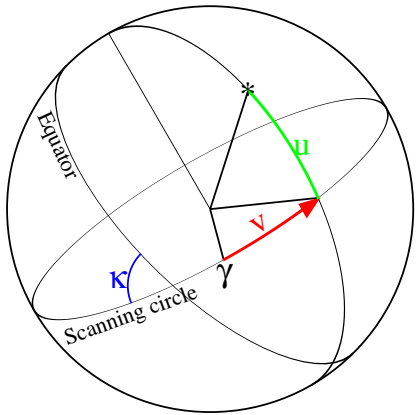}{5cm}{0}{30}{30}{-80}{-60}
\caption{\label{Fig:scan}All what Hipparcos measured are the angles $v$, the abscissae, along a precessing scanning great circle.  The precise knowledge of the satellite attitude yields $\kappa$.}
\end{figure}

For each observed star, Hipparcos (ESA 1997) measured tens of abscissae, i.e. 1-dimensional quantities, along a precessing great circle (Fig.~\ref{Fig:scan}).
Several corrections were applied to these abscissae (chromaticity effects, orientation errors, \dots).  Different models were then used to fit these abscissae, ranging from a single star model with just five parameters to the orbital model with up to twelve parameters fitted (the DMSA/O part of the Hipparcos catalogue lists all these orbital solutions).

Together with the catalogue, it was decided that the residuals of these corrected abscissae would be released also.  In order to make the interpretation of these residuals unique, the values released were all derived with the single star model, whether it was used for that entry in the catalogue or not.  That makes it possible for anybody to fit any model thus looking for further reduction of the residuals.  The fit is achieved through a $\chi^2$ minimization.  For instance, in the case of an orbital model, its expression looks like:
\begin{equation}
\chi^2=(\Delta v-\sum_p\frac{\partial v}{\partial p}\Delta p-\sum_o\frac{\partial v}{\partial o}o)^{\mbox{t}}\vec{V}^{-1}(\Delta v-\sum_p\frac{\partial v}{\partial p}\Delta p-\sum_o\frac{\partial v}{\partial o}o) \label{Eq:chi2}
\end{equation}
where $\Delta p$ is the correction with respect to the original parameter $p$ and $\vec{V}$ is the covariance matrix of the data.   $\Delta v_j, \partial v_j/\partial p_k,$ and $V$ ($j=1,\dots, n; k=1, \dots, 5$) and the Main Hipparcos solution are provided (see van Leeuwen \& Evans (1998) for details).  The partial derivative of $v$ with respect to the orbital parameter $o$ is given by
\[
{\partial v \over\partial o}={\partial v\over\partial p_1}{\partial \xi\over \partial o}+{\partial v\over\partial p_2}{\partial \eta\over \partial o}
\]
where $(\xi, \eta)$ are the astrometric coordinates of the photocenter in the plane orthogonal to the line of sight.  One simply needs to express $o$ in terms of the orbital parameters and then minimizes Eq.~(\ref{Eq:chi2}).  

Depending on the number of parameters assumed from a ground based solution, finding the minimum of this expression can be very tough.  Indeed, if $(e,P,T)$ are assumed, the model is linear in terms of the other nine parameters and the above chi square hence has a unique minimizer.  Any optimization algorithm will find it.  Together with the case where the semi-major axis of the orbit is the only parameter left free, these are the two situations where the model is linear.  In any other situation, it is important to carefully handle the many local minima of the chi square in order to identify the deepest one (Horst et al. 1995).

For the sake of completeness, one should stress that besides the IAD, there are also the Transit Data (TD), a byproduct of the processing of the binaries by the Northern Data Analysis Consortium.  However, TD are available for some Hipparcos entries only.  Quist \& Lindegren (1999) give a detailed description of the TD as well as on how to use them. 

\section{Improving the parallax/proper motion}

Even in cases where the orbital parameters are not the primary target of the fit, there are some situations in which it is important to account for the duplicity through the use of the orbital model.  
\begin{itemize}

\item When $P\sim 1$ year, there is a strong correlation between $\varpi$ (the parallax) and the size of the orbit (Pourbaix \& Jorissen 2000).  Indeed, the effect of the parallax on the position of the star has a period of one year.  Therefore, any other phenomenon with the same period (e.g. the orbital motion) will be indistinguishable from the parallactic motion.  Depending on the orientation of the orbit, the resulting parallax can be enlarged or shrunk with respect to the truth.  Ignoring the orbital motion here thus affects the {\em accuracy} of the fitted parallax.

\item A short orbital period shows up as an additional noise on the observations.  When the orbital period is much shorter than one year, no confusion is possible between the two motions and the derived parallax is not affected.  However, when the orbital motion is not accounted for, the observations exhibit some extra scatter which affects the {\em precision} of the whole single star solution.

\item When the orbital period is much longer than the mission duration, the motion of the star on its orbit is just a small arc or even a segment of line.  This can lead to a confusion between the orbital motion and the proper motion.  There are 2\,622 Hipparcos entries (DMSA/G) for which an acceleration term in the position or proper motion was fitted: up to four additional parameters were adjusted together with the basic five astrometric parameters.  The accuracy of the proper motion is particularly valuable for those interested in galactic dynamics or cluster membership. 

\item An orbit was sometime imposed prior to the reduction, with no orbital fit.  In four cases, namely $\mu$ Cas, $\epsilon$ Aur, Procyon and Sirius, the orbital model was used in order to lower the risk of grid-step error but no orbital parameter was fitted.  The resulting fundamental astrometric parameters are thus tied to the adopted orbit.  When the latter is refined, the position, parallax and proper motion should be revised accordingly (e.g. Procyon, Girard et al. 2000).
\end{itemize}

These effects on the parallax and proper motion for long period binaries are illustrated by Pourbaix \& Jorissen (2000) with lately discovered binaries among Ba, CH and Tc-poor S stars.

\section{The inclination: the holy grail}\label{Sect:Grail}

The orbital model improves the accuracy and precision of the five basic astrometric parameters but the main reason for modeling the observations that way is nevertheless to derive the orbital inclination, unaccessible to the radial velocity techniques.

The equations are essentially the same as for visual binaries, one simply replaces the semi-major axis of the relative apparent orbit with that of the absolute photocentric one.  However, there is a major difference between the two situations.  In the case of a visual orbit, one sees the two components and one waits long enough for an orbital arc to show up.  Here, there is a companion but one does not know whether the amplitude of the resulting astrometric wobble is large enough with respect to the satellite precision or not.  As we will see, fitting the noise with an orbital model can have some awful consequences.

\subsection{Stellar applications: episode 1}
Besides the original Hipparcos catalogue, the very first applications of its observations to stellar orbits are those of Halbwachs et al. (2000) and Pourbaix \& Jorissen (2000).  Whereas van Leeuwen \& Evans (1998) present the data and how to use them, they do not apply them to binaries.

Halbwachs et al. (2000) investigate eleven spectroscopic binaries with brown dwarf candidates, looking for the actual mass of the companions.  By fitting a population of genuine single stars with an orbital model, they guess how overestimated the semi-major axis of the orbits of genuine binaries is.  Correcting for that effect, they conclude that only three stars of their sample are viable brown dwarf candidates.

The methodology of Pourbaix \& Jorissen (2000) is rather different.  They study a sample of 81 chemically-peculiar red giants (see the paper by North \& Debernardi in this volume) belonging to single-lined systems.  They also assume that the unseen companions are white dwarfs and thus fix their mass to the average mass of field WD's.  The IAD (and sometime TD) are fitted and three (respectively four) orbits are derived for each system.  They are based respectively on the FAST data, NDAC data and FAST+NDAC data (the fourth orbit being derived from the TD).  The point of the authors is that if the observations exhibit the astrometric wobble caused by the secondary, the three (four) solutions should agree.  This assessment criterion together with others based on some astrophysical considerations lead the authors to keep 23 systems only.

Gatewood et al. (2000) combined the IAD of the triple system $\pi$ Cep together with a spectroscopic orbit and some MAP (Gatewood 1987) observations.  They thus obtained the individual mass of the three components of this system.  In that particular case, the agreement between the MAP and Hipparcos results was used to assess their likelihood.

\subsection{Extrasolar planet furry}

If it works for binary systems, it should work for extrasolar companions too!  Well, indeed, the equations are the same in both cases.  That is however where the comparison stops making sense.

The first application of the Hipparcos data to extrasolar planets was by Perryman et al. (1996).  Using the spectroscopic elements of three newly discovered planets, the authors derived an upper limit on the mass of the companion.  Even though the authors could not confirm these objects are planets, they nevertheless concluded upon their substellar nature.

Whereas Perryman et al. (1996) used the absence of orbital signal in the data to set an upper bound on the companion mass, Mazeh et al. (1999) fitted the IAD of $\nu$ And in order to derive the mass of its outer planetary companion.  Although their result is below 13 Jupiter masses, a $1\sigma$ error would already be above that threshold.  Zucker and Mazeh (2000) applied the same technique and concluded upon the brown dwarf nature of the companion of HD~10697.

The technique being {\em so promising} for the extrasolar planet field, Han et al. (2001) applied it to 30 systems with a candidate planetary companion with a period larger than 10 days.  Their result presented at a press conference had the effect of an Earthquake: most of the inclinations were quite low, resulting in stellar rather than planetary companions.

Whereas one cannot rule out one nearly face-on orbit, the random orientation of the orbital planes, i.e. a uniform distribution of $\sin i$, favors edge-on like orbits.  Han et al. therefore suggested an observational bias towards small inclinations.  That was simply too much!  

\subsection{Screening methods}

Though Halbwachs et al. (2000) already considered the case of low S/N data, they mainly focused on the effect on the semimajor axis.  Pourbaix (2001) showed that even in absence of signal, the method adopted by Han et al. (2001) yields low inclinations.  In case of pure noise, it leads to face-on orbits with very small `error bars'.  Thus, though it was not a proof that the result by Han et al. was wrong, it was nevertheless a proof that one would obtain the very same result even if Hipparcos had not noticed the astrometric wobbles caused by the companions.

\begin{figure}[htb]
\plotfiddle{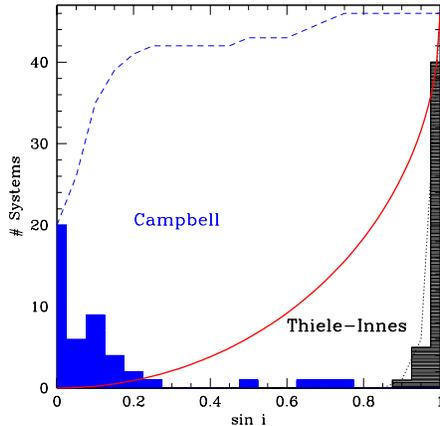}{5cm}{0}{30}{30}{-80}{-60}
\caption{\label{Fig:distrInc}Distribution of the orbital inclination after two distinct approaches.  The thick line is the theoretical cumulative distribution corresponding to a random orientation.  The dashed-line (resp. dotted-line) is the cumulative distribution of the derived inclinations according to Campbell's (resp. Thiele-Innes') method.}
\end{figure}

Pourbaix \& Arenou (2001) went further and concluded that the size of these wobbles was way below the Hipparcos detection limit.  Their conclusion is based upon the fact that if the observations do exhibit the orbital signal, two mathematically equivalent methods will lead to the very same result (within the error bars).  The two methods assume $e$, $P$ and $T$ from the spectroscopic orbit.  In one case (Campbell's method), $K_1$ and $\omega$ are also adopted thus leading to only two fitted parameters, namely $i$ and $\Omega$.  In the other method, the four Thiele-Innes constants are fitted thus allowing the Hipparcos based $\omega$ to be compared with the spectroscopic one.  The distributions of the inclinations after the two methods are plotted in Fig.~\ref{Fig:distrInc}.  Even if the Thiele-Innes based inclinations are more consistent with the expectation, they are as spurious as the Campbell's ones: both are model artifacts.

In order to quantify the likelihood of the fit of an astrometric orbit, Pourbaix \& Arenou introduced several statistical indicators whose combination allows to keep or discard such a solution.  Let us list a few of them.
\begin{itemize}
\item The addition of 4 supplementary parameters (the four Thiele-Innes orbital constants) describing the orbital motion should
result in a statistically significant decrease of the $\chi^2$ for the fit of the $N$ IAD with an
orbital model ($\chi^2_T$), as compared to a fit with a single-star solution ($\chi^2_S$). This criterion is expressed by an $F$-test:
\begin{equation}
Pr_2 = Pr(\hat{F} > F(4,N-9)),\;\; \mathrm{where} \;\; \hat{F} =  \frac{N-9}{4}\;\; \frac{\chi^2_S - \chi^2_T}{\chi^2_T}.
\end{equation}
$Pr_2$ is thus the first kind risk associated with the rejection of the null hypothesis:``{\em there is no orbital wobble present in the data}''.

The advantage of the F-test over a simple $\chi^2$-test is that the former does not rely upon the weights of the observations or their quoted uncertainties.  Indeed, two models computed with the same weights are derived and compared.  Thus, if all the uncertainties are underestimated by a factor 10, the chi square will be overestimated by a factor 100 whereas the F-test will not be affected at all.

\item Getting a substantial reduction of the $\chi^2$ with the Thiele-Innes model does not necessarily imply that the four
Thiele-Innes constants $A,B,F,G$ are significantly different from 0. The first kind risk associated with the rejection
of the null hypothesis ``{\em the orbital semi-major axis is equal to zero}''
may be expressed as
\begin{equation}
Pr_3 = Pr(\chi^2_{ABFG} > \chi^2(4)),\;\; \mathrm{where}\;\; \chi^2_{ABFG} =  \vec{X}^t \vec{V}^{-1} \vec{X},
\end{equation}
and $\vec{X}$ is the vector of components $A,B,F,G$ and $\vec{V}$ is its covariance matrix.  With the above notations, the requirements for a star to qualify as a binary may then be encapsulated at once in the expression
\begin{equation}
\label{Eq:alpha}
\alpha \equiv (Pr_2 + Pr_3)/\epsilon \le 0.2.
\end{equation}

\item For the orbital solution to be   a significant one, its parameters should not be strongly correlated with the  other astrometric parameters (like, e.g., the proper motion). In other words, the covariance matrix of the
astrometric solution should be dominated by its diagonal terms, as measured by the {\it efficiency} $\epsilon$ of the matrix being close to 1 (Eichhorn 1989). The efficiency is simply expressed by 
\begin{equation}
\epsilon = \sqrt[m]{\frac{\Pi_{k=1}^m \lambda_k}{\Pi_{k=1}^m \vec{V}_{kk}}},
\end{equation}
where $\lambda_k$ and $\vec{V}_{kk}$ are respectively the eigenvalues and the diagonal terms of the covariance matrix $\vec{V}$.
\end{itemize}

Zucker and Mazeh (2001) proposed a completely different approach for assessing these astrometric orbits.  Their method is based on a permutation test.  The timing of all IAD are permuted and the partial derivatives $\partial v/\partial p$ updated accordingly.  This new set of observations is then fitted with an orbital model.  The semimajor axis obtained with the original dataset is compared to the ones based on the new datasets.  The assessment is based on the percentage of resulting semimajor axes below the original one.  The higher the percentage, the more reliable the original fit.  That method led Zucker and Mazeh to reject their own previous results on $\nu$ And and HD~10697.

\subsection{[Sub]stellar applications: episode 2}

The criteria adopted by Pourbaix \& Arenou (2001) have been extended by Pourbaix \& Boffin (2003).  For instance, the latter systematically carry on a period analysis in order to see whether the Hipparcos data exhibit a peak at the spectroscopic period.  The size of the orbit is also compared with the Hipparcos precision.  For instance, systems with $a_0\sin i<1.5$ mas are immediately discarded.  With this toolbox available, it is now possible to tackle even large samples of binaries in a very automated way with no assumption about any of the component of the system (as Pourbaix \& Jorissen (2000) did).  That is exactly what is needed for large missions like Gaia.

Pourbaix \& Boffin (2003) applied these tools to a sample of systems with a giant component whereas Pourbaix et al. (2004) apply them to all the single-lined binaries listed in the 9th catalogue of spectroscopic binary orbits.\footnote{http://sb9.astro.ulb.ac.be}  As shown by Pourbaix \& Boffin, the reliability of some spectroscopic orbits is likely to be the reason for rejecting the corresponding astrometric solutions (e.g.  HIP~29982).

Some recent papers on extrasolar planets (Jones et al. 2002, Marcy et al. 2002, Vogt et al. 2002, Fischer et al. 2003) made use of these tools to check whether Hipparcos can give some clue on the inclination or not.  Even though the conclusion was always negative, it is worth stating it immediately.  Frink et al. (2002) used the IAD to set an upper bound on the mass of a substellar companion to a giant star.

\section{Searching for binaries}\label{Sect:Searching}

As shown by Pourbaix (2002), 1-dimensional data have their own limitations which have almost nothing to do with their actual precision or even with the fraction of the orbit covered by the mission.  The distinction should be made clear between detection and orbit determination.  The former can be achieved with a simple $\chi^2$-test.  However, that the test states the object is not single does not mean one can derive its orbit from scratch.

Besides deriving orbits of known systems, the Hipparcos data can also be used to identify new binaries.  That may be very useful if, for instance, one wants to estimate the binary proportion within certain star samples.  Two approaches have been suggested: one relying upon the Hipparcos data only and another based on the comparison between the Hipparcos results and those from other catalogues.

The Tycho-2 catalogue (H\o g et al. 2000) is an extension of the Tycho catalogue (ESA 1997) where the proper motion does no longer rely on the sole Tycho observations.  That proper motion is thus much more accurate than the Tycho and Hipparcos ones.  It is also free of orbital contribution even for long period binaries.  Makarov (2004) lately derived the orbit of van Maanen 2 using the Hipparcos data and proper motion after Tycho-2.  The two proper motions are discrepant thus indicating that the Hipparcos one is affected by the orbital motion whereas the Tycho-2 is not.  Even if one cannot always be as lucky as Makarov and derive an orbit, one can nevertheless undertake such comparisons and expect to identify some new binaries.

Another approach for searching for binaries is to rely on the Hipparcos data only, using the screening tests described in Sect.~4.3 (especially Eq.~\ref{Eq:alpha}).  Hipparcos data are, however, seldom precise enough to derive the orbital elements from scratch. Therefore, when a spectroscopic orbit is available beforehand, it is advantageous to import $e, P, T$ from the spectroscopic orbit and to derive the remaining astrometric elements (Pourbaix \& Boffin 2003).  If a spectroscopic orbit is not available, trial $(e, P, T)$ triplets scanning a regular grid (with $10 \le P (\rm d) \le 5000$ imposed by the Hipparcos scanning law and the mission duration) may be used. The quality factor $\alpha$ is then computed for each trial $(e, P, T)$ triplet, and if more than 10\% of these triplets yield $\alpha < 0.2$, the star is flagged as a binary. This approach is applied by Jorissen et al. (this volume) on a sample of barium stars. These authors show that, in most cases, this method makes it even possible to find a good estimate for the orbital period, provided, however, that the true period is not an integer fraction, or a multiple, of one year.

\section{Conclusion and perspectives}

Even though Hipparcos observed for three years only and its observations are 1-dimensional, they are precise enough to allow for some orbit fitting, either from scratch or with the help of some ground-based data (e.g. spectroscopic orbit, \dots).  The Hipparcos catalogue is the best result that could be achieved within the time constraints from ESA.  It is therefore not surprising that some results can be improved (e.g. parallax) or derived for the first time (e.g. orbital inclination).  It is sometime just a question of time we are willing to spend on a specific object.

Owing to the similarity of its observations with those of Gaia, the Hipparcos data are an excellent testbed for the Gaia data reduction pipeline.  Statistical tests (Pourbaix \& Arenou 2001) can be assessed and made fully ready for Gaia.  Though Gaia will supersede Hipparcos, the data of the latter have not yet revealed all their secrets.  So, their usefulness is double: derive some scientific results further on and get prepared for the next mission.

\acknowledgments This research was supported in part by an ESA/PRO\-DEX Research Grant.

%\end{document}

\clearpage
\newpage

\setcounter{section}{0}
\setcounter{subsection}{0}
\setcounter{subsubsection}{0}
\setcounter{table}{0}
\setcounter{figure}{0}

\markboth{Jorissen et al.}{Binaries in the Hipparcos Data: Keep digging I.}
\pagestyle{myheadings}
%\nofiles

%\marginparwidth 1.25in
%\marginparsep .125in
%\marginparpush .25in
%\reversemarginpar

%\begin{document}
\title{Binaries in the Hipparcos data: Keep digging\\
{\small\bf I. Search for binaries without {\it a priori}
knowledge of their orbital elements: Application to barium stars}}
\author{A. Jorissen, S. Jancart, D. Pourbaix} 
\affil{Institut d'Astronomie et d'Astrophysique, Universit\'e Libre de Bruxelles CP 226, 
Boulevard du Triomphe, B-1050 Bruxelles, Belgium} 

\begin{abstract}
This work makes use of {\it Hipparcos} data to test the algorithms of (i)
binary detection and (ii) orbital-parameters determination, which could
possibly be used in the GAIA pipeline. The first item is addressed in
this paper, whereas the second one is addressed in a companion paper
by Pourbaix et al. (this volume). Here we test the ability of the
algorithm to detect binaries from scratch, {\it i.e.,} from the astrometric
data without any {\it a priori} knowledge of the orbital elements. The
Hipparcos {\it Intermediate Astrometric Data} of a complete sample of
163 barium stars (supposed to be all members of binary systems from
theoretical considerations) are used as test bench.  
When $\varpi > 5$~mas and  $P < 4000$~d, 
the binary detection rate is close to 100\%, but it falls to 22\% when considering the 
whole sample, because many barium stars have small parallaxes or very long periods. 
\end{abstract}

The algorithm to detect
binaries using astrometric data only, as described in Sects.~4.3 and 5
of Pourbaix (this volume) has been applied to the existing Hipparcos 
{\it Intermediate Astrometric Data}
(hereafter IAD; van Leeuwen \& Evans 1998) of barium stars.
Barium stars constitute an ideal sample to test this algorithm, because
they are all members of binary systems (Jorissen et al. 1998), with periods ranging 
from about 
100~d to more than 6000~d. The catalogue of L\"u et al. (1983) contains  163 {\it bona fide}
barium stars with an HIP entry
(excluding the supergiants HD~65699 and HD~206778 = $\epsilon$~Peg).

When  
$\varpi > 5$~mas and  $P < 4000$~d (upper left corner of Fig.~1, left panel), 
the (astrometric) binary detection rate is close to 100\%, but it falls to 22\% (=36/163) 
when considering the 
whole sample, because many barium stars have small parallaxes or very long periods. 
Astrometric orbits with $P > 4000$~d (11~y)
can generally not be extracted from the Hipparcos IAD, which span only 3~y 
(see left panel of Fig.~1).
Similarly, when $\varpi < 5$~mas, as have most barium stars  (right panel of Fig.~1), 
the Hipparcos IAD  are not precise enough to extract
the  orbital motion.

An interesting astrophysical outcome of the present
work is the list of barium stars shown to be astrometric 
binaries by the analysis of the Hipparcos IAD. They are listed in Table~1, with an estimate 
of their orbital period, as obtained from Fig.~2.

%Because parallactic motion has 1-y period, 1-year aliases are strong, and prevent finding reliable orbital solutions
%with periods close to multiple or integer fractions of 1~y (see Fig.~3). Inspection of the run of $\alpha$ versus
%the trial periods $P$ (Fig.~3) makes it even
%{green}{possible in many cases to guess the most probable orbital period}, as discussed in the next section
%(see also Table~1).

{\footnotesize {\bf Acknowledgments.} A.~J. and D.~P. are Research
Associates, F.N.R.S (Belgium). Financial support to this work was
provided through ESA/PRODEX Grants
90078 and 15152.}

\begin{table}
\caption{Ba stars with definite evidence for astrometric binary motion. 
%Whenever possible, the orbital period has been estimated (see Fig.~2). 
}
\begin{tabular}{rrlr|rrlr}
\tableline
HIP   & \multicolumn{1}{c}{$\varpi$}  & Sp. Typ. &
\multicolumn{1}{c}{$P$} & \multicolumn{1}{c}{HIP}   &
\multicolumn{1}{c}{$\varpi$}  & Sp. Typ. & \multicolumn{1}{c}{$P$} \\
       & (mas)    &          & \multicolumn{1}{c}{(d)}
&        & (mas)    &          & \multicolumn{1}{c}{(d)}\\
\tableline
944   &  7.11 & K0p & $>800$     & 89386 & 5.69  & K1 Ba1 & $>500$ \\
10119 &  8.91 & F6 Ba1$^\dagger$ & $>1000$ & 89881 & 9.37  & G9 Ba1 & 900\\
21681 &  3.44 & K1 Ba1 & $>1000$ & 90316  & 0.56  & K0 Ba1 & 500\\
25547 &  3.43 & K1 Ba1 & $>1000$ & 93188  & 5.55  & K3 Ba1 & 200?\\
38488 &  1.29 & K2 Ba2 & $>1000$ & 96024  & 3.48  & K0 Ba2 & $>800$\\
47267 &  6.73 & K0 Ba1 & $>1000$ & 97613  & 4.10  & K1 Ba2 & ?\\
47881 &  3.56 & K0 Ba1 & $>1000$ & 105294 & 3.49  & F2 Ba1$^\dagger$ & ? \\
53091 &  2.14 & K1 Ba1 & 400 & 107685     & 1.77  & K2 Ba2 & 700\\
65535 &  15.73 & K1 Ba3 & $>800$ & 117585 & 7.08  & K2 Ba2 & ?\\
\tableline
\tableline
\noalign{$^\dagger$dwarf Ba star} 
\end{tabular}
\end{table}
\begin{figure}
\plottwo{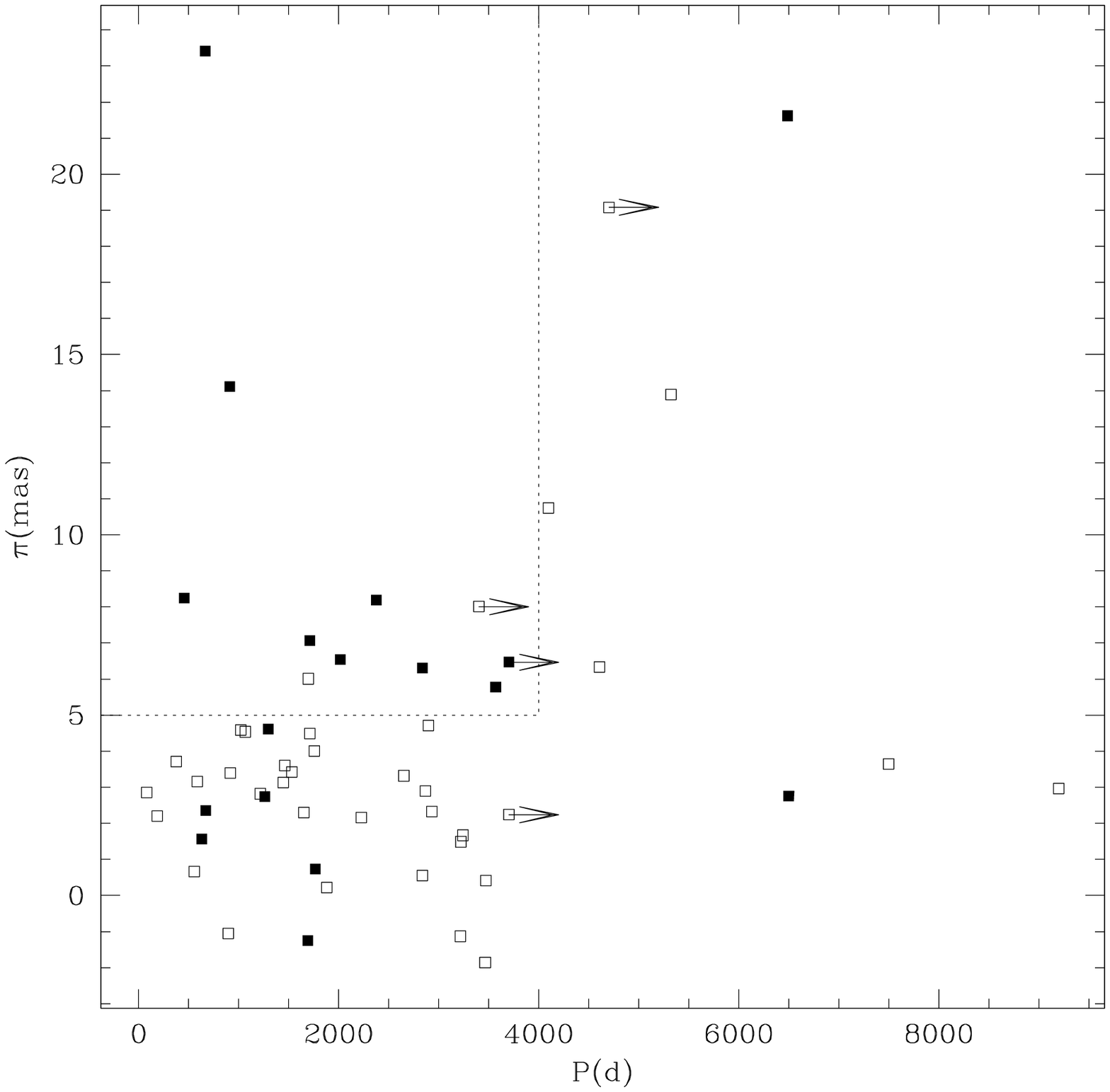}{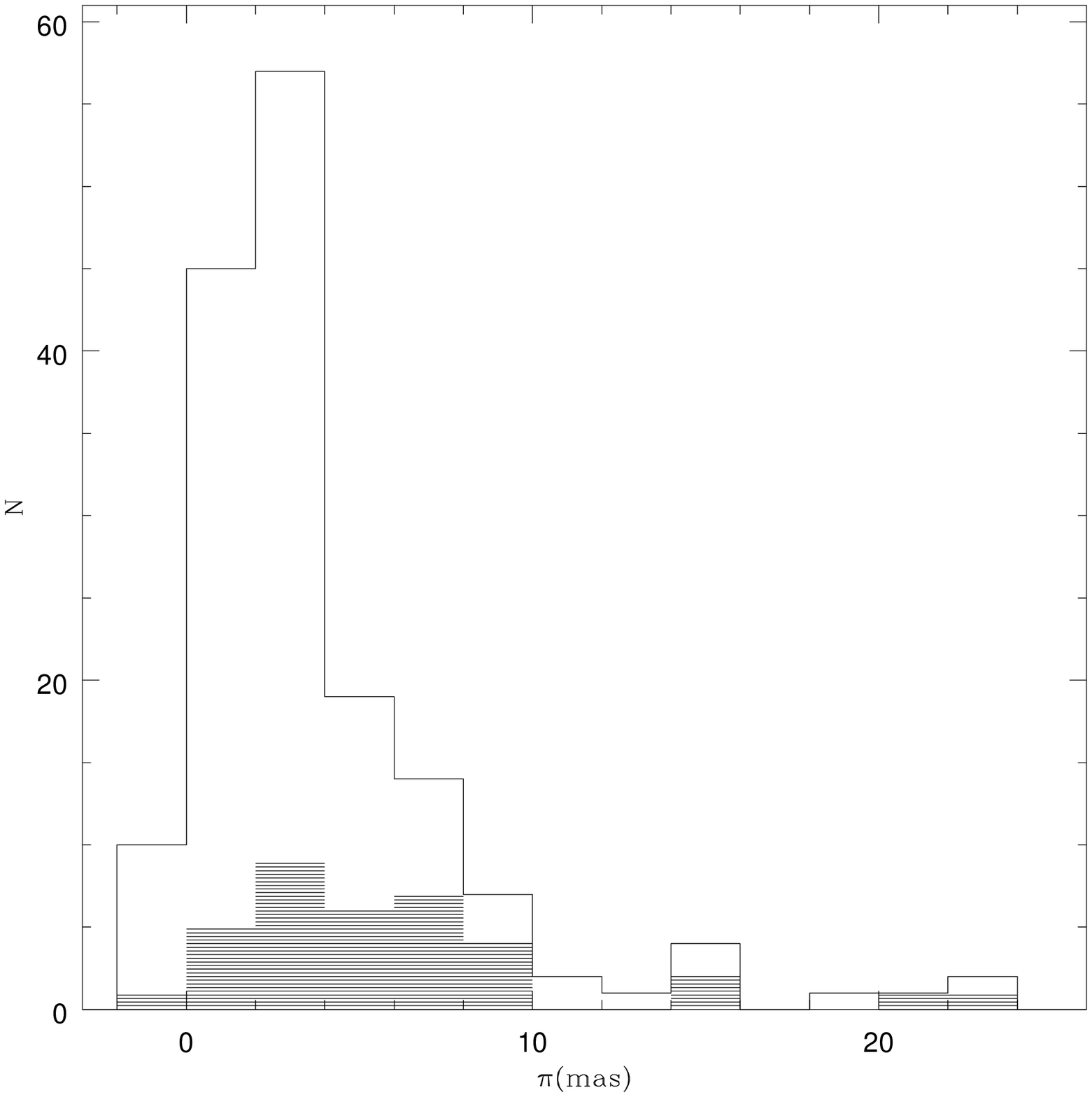}
\caption{{\bf Left panel:} 
Barium stars (previously known as SBs) 
flagged as astrometric binaries by the algorithm are represented by
  black symbols.  
Arrows denote stars with only a lower limit available on the period. 
{\bf Right panel:} Fraction of stars flagged as binaries (shaded
histogram) compared to 
total number of stars, as a function
of parallax.  
As expected, the detection rate becomes high for $\varpi > 5$~mas.
}
\end{figure}

\begin{figure}
\plottwo{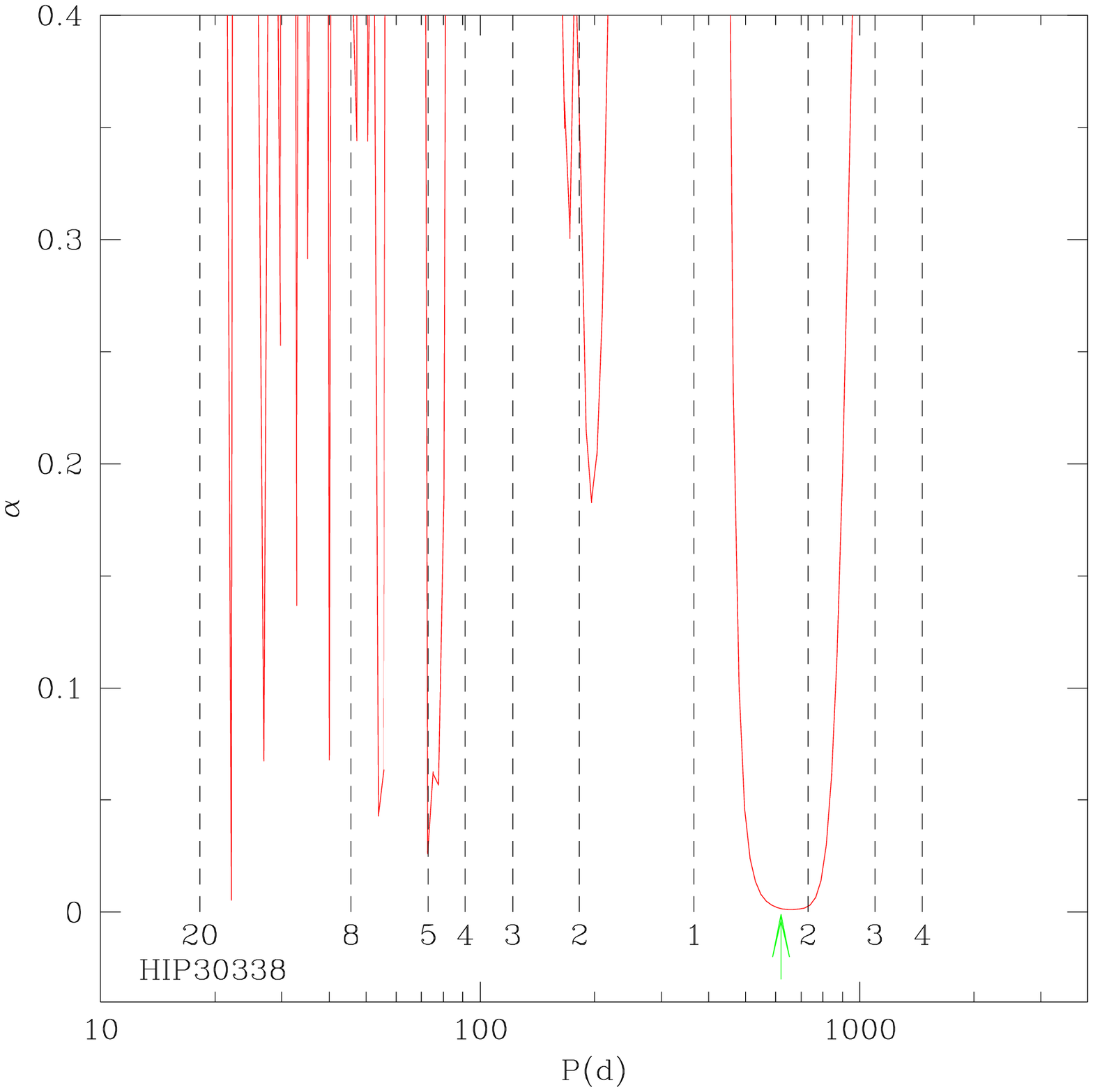}{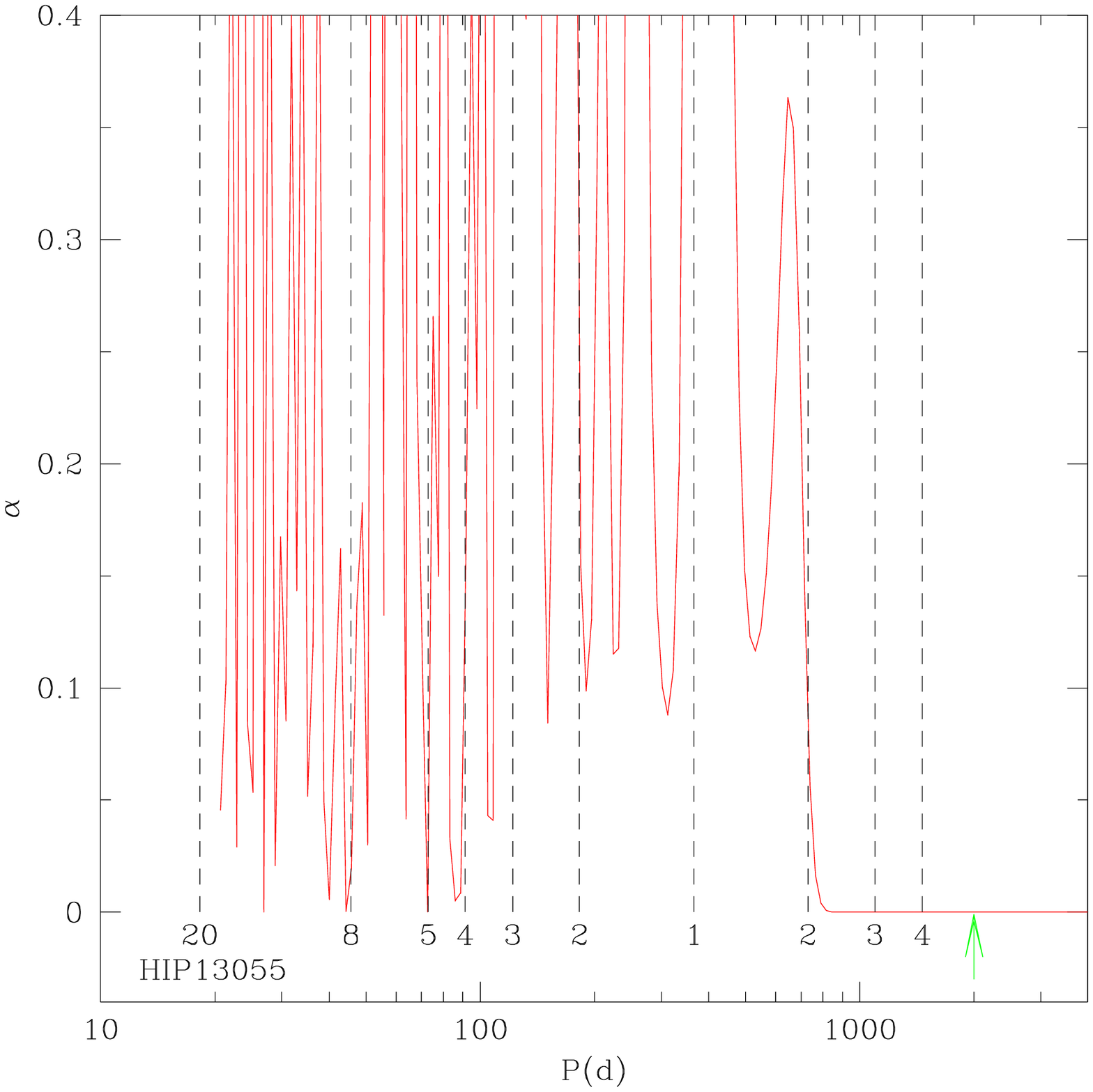}
\mbox{}\\
\plottwo{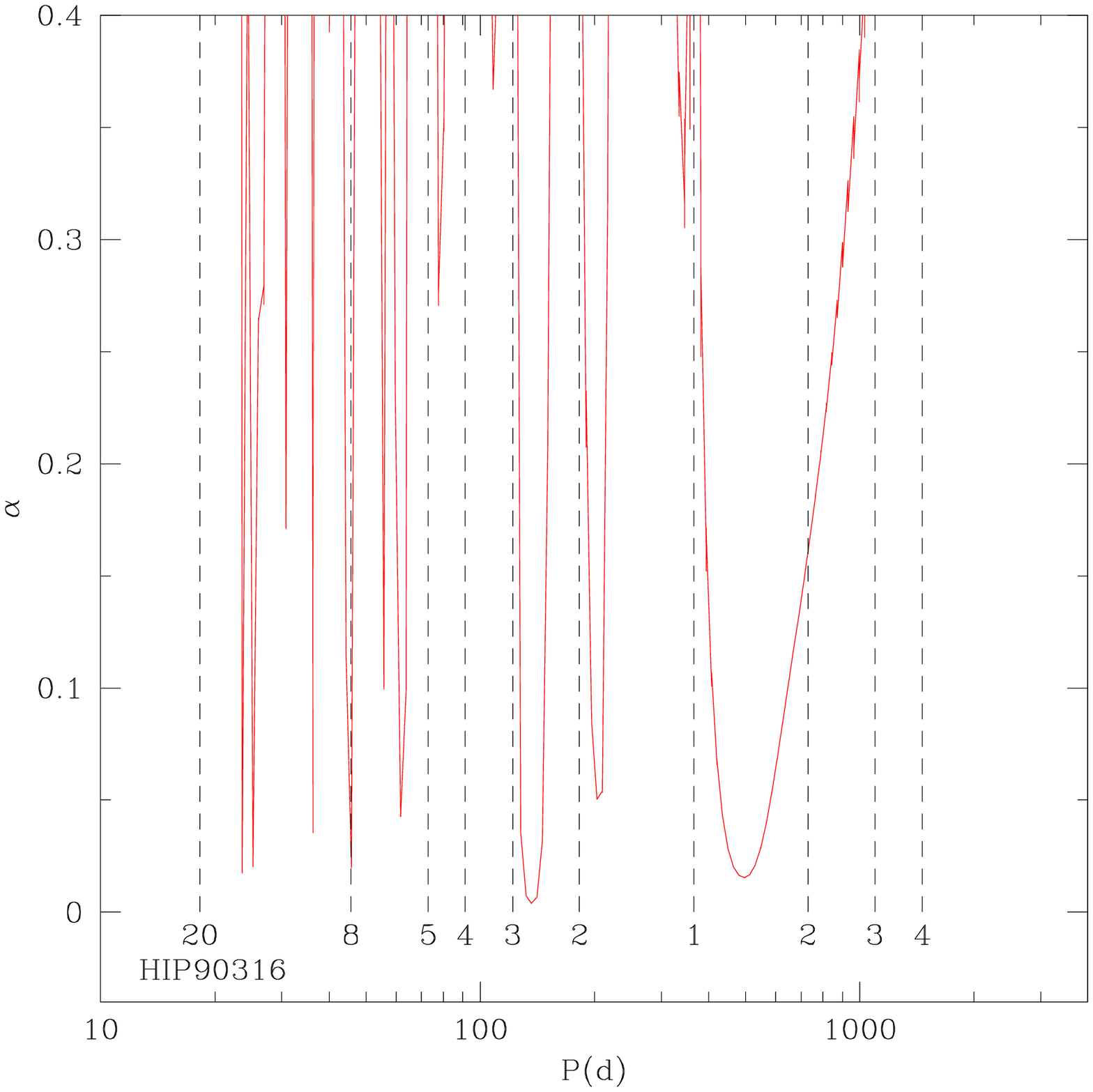}{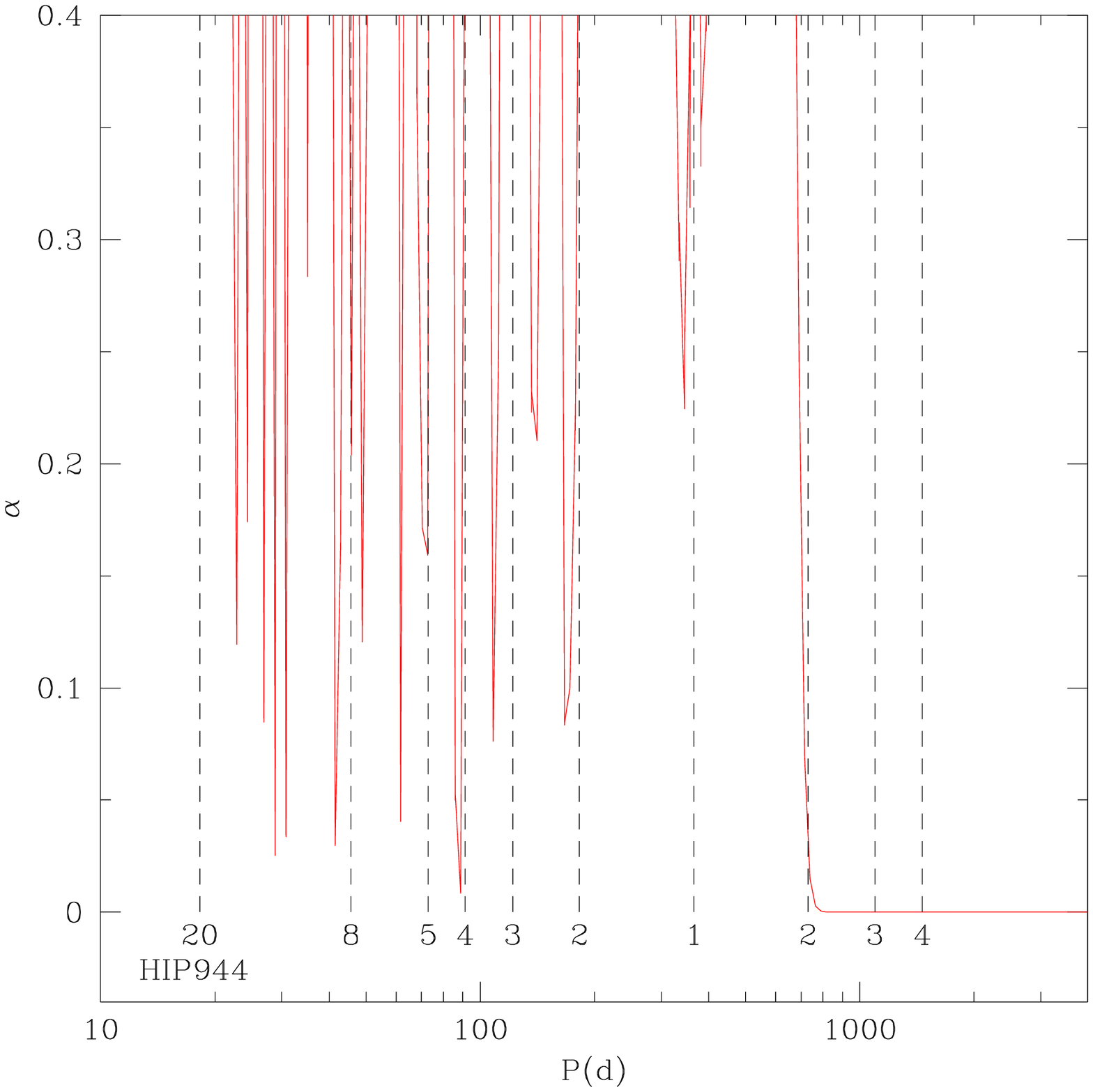}
\caption{The $\alpha$ statistics (see Eq.~5 of Pourbaix, this volume) 
as a function of the trial orbital period (assuming $e
= 0$) for confirmed spectroscopic binaries (upper row), and for stars newly flagged as binaries by 
the present
algorithm (lower row). In the former case (upper row), the known spectroscopic period 
(represented by an arrow)
indeed lies within the range of minimum 
$\alpha$ values. By comparison, the new binaries 
HIP~944 and 90316 are likely to have periods $P > 800$~d and $\sim 500$~d, 
respectively. 
The vertical dashed lines represent multiple, or integer fractions, of 1~y. At those periods, there 
is a strong correlation between the parallactic and orbital signals, which degrades the $\alpha$ statistics
and makes binaries difficult to find at those 1-y alias periods.} 
\end{figure}

\clearpage
\newpage

\markboth{Pourbaix et al.}{Binaries in the Hipparcos data: Keep digging II}
\pagestyle{myheadings}
\def\SB9{$S _{B^9}$}

%\begin{document}

\setcounter{section}{0}
\setcounter{subsection}{0}
\setcounter{subsubsection}{0}
\setcounter{table}{0}
\setcounter{figure}{0}

\title{Binaries in the Hipparcos data: keep digging\\ {\small \bf II. Modeling the IAD of known spectroscopic systems}}
\author{D.~Pourbaix, S.~Jancart, A.~Jorissen}
\affil{Institute of Astronomy and Astrophysics, Universit\'e Libre de Bruxelles CP 226, Bld du Triomphe, B-1050 Bruxelles}

\section{Astrometric/Spectroscopic combination}

It is well known that the Hipparcos (ESA 1997) satellite was originally designed to measure positions, proper motions and parallaxes of stars.  However, the combination of the Hipparcos Intermediate Astrometric Data (IAD) with ground-based spectroscopic data have lately led to nice results about specific binaries and extrasolar planets. For instance, the IAD of known spectroscopic binaries can be fitted with an (astrometric) orbital model thus hopefully yielding the inclination of the system.  The key point is to understand the limitation of the method.

The spectroscopic orbits were all extracted from the 9th catalogue of spectroscopic binary orbits (hereafter \SB9, {\em http//sb9.astro.ulb.ac.be}).  Even though \SB9\ does contain 1\,578 systems, we only kept the 863 which do not belong to the DMSA/C (ESA 1997).  When several subsystems for a given HIP number were in \SB9, only the one with the largest orbital period was kept.  

130 systems simultaneously fulfill the six criteria after Pourbaix \& Boffin (2003) are used.  We adopt the same definition as given by Pourbaix in this volume and impose: $pr_2, pr_3 \le 5 \%$ (see Pourbaix \& Arenou (2001) for details), the efficiency $\epsilon \ge 0.4$ (Eichhorn 1989), the consistency between the Thiele-Innes solution and the Campbell/spectroscopic one ($pr_4 \ge 5 \%$), the likelihood of the face-one orbit ($pr_5 \le 5 \%$) and the consistency of the spectroscopic and astrometric orbits ($|D| < 2$).

\section{Periodogram analysis}

\begin{figure}[t]
\plotfiddle{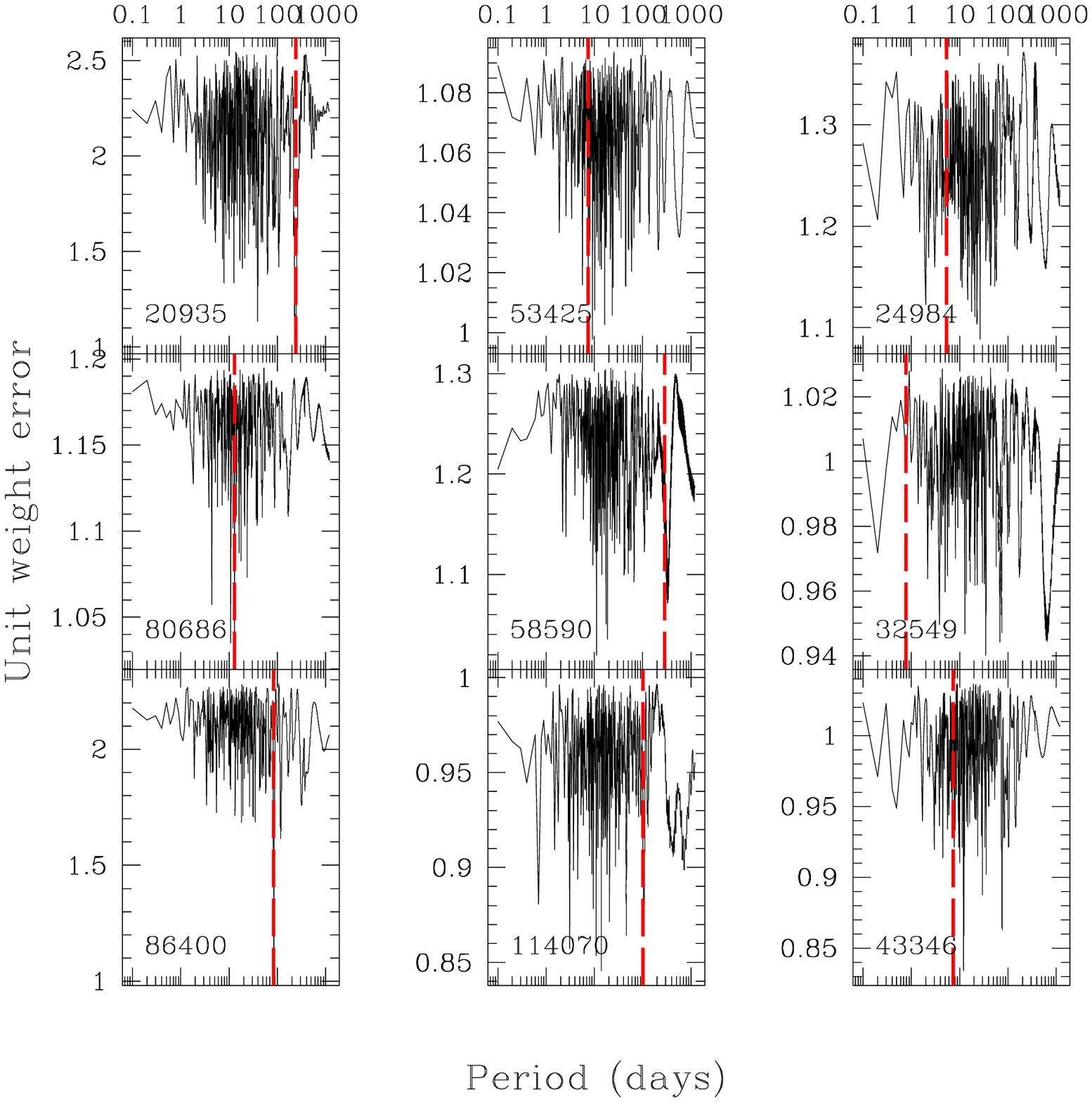}{5cm}{0}{30}{30}{10}{-60}\hfill
\plotfiddle{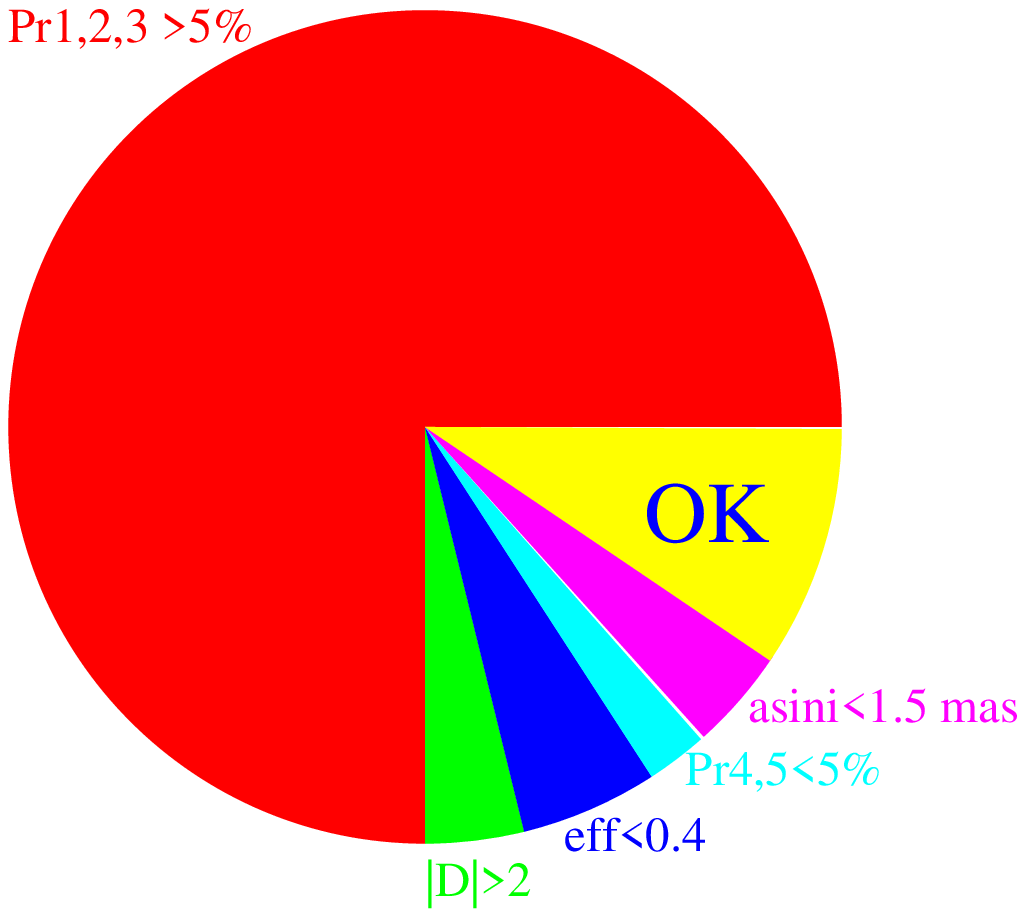}{1cm}{0}{60}{60}{-60}{-60}
\caption{\label{Fig:9periods}{\bf Left panel:} Periodogram and the orbital period (dashed lines) for typical cases. Left column: The strongest peak in the periodogram matches the spectroscopic period. Central column:  The period and one peak are close.  Right column: No peak corresponds to the spectroscopic period. {\bf Right panel:} Reasons for discarding systems}
\end{figure}

All these tests are based on the fit.  Hence, a positive value does not mean that what Hipparcos saw is the companion we are looking for.  A periodogram is used to assess the match between these systems: do the IADs contain a peak at the period corresponding to the spectroscopic one (dashed line on figure~\ref{Fig:9periods})?  It is worth keeping in mind that these periodograms rely on the eccentricity and periastron time of the adopted spectroscopic orbit.

Sixty eight systems exhibit a peak in the periodogram corresponding exactly to the spectroscopic period (left column of the left panel of figure~\ref{Fig:9periods}).  For 36 systems, the periods are closed or may correspond to one peak but not the one in evidence in the periodogram (central column of the left panel of figure~\ref{Fig:9periods}).  Finally, 19 systems end up with a spectroscopic period completely out of a peak (right column of figure~\ref{Fig:9periods}).  Seven systems have orbital periods exceeding the limit of the periodogram.

\section{\SB9\ and BDB informations}
In order to understand the reason of the discrepancy noticed for some systems, we used informations from \SB9 and from the Besan\c{c}on Double and Multiple Star Data Base (BDB).  Several explanations are necessary:
\begin{itemize}
\item Right spectroscopic period, but
\begin{itemize}
\item Sampling problem: when generating the periodogram, the closest period to the true one is still too far away
\item scanning law constraint: the Hipparcos scanning law prevents some periods from being clearly identified.
\end{itemize}
\item Wrong spectroscopic period
\begin{itemize}
\item The spectroscopic period of a few stars are flagged as uncertain or preliminary by the authors.
\item Some are based on very old observations, beginning of last century and have not been checked for the past fifty years.
\item In \SB9 some binaries we are interested in are given with a poor grade of reliability.
\end{itemize}
\item Unaccounted triple systems: even though the system does not belong to the DMSA/C and does not appear as a triple in \SB9, a long period shows up in the periodogram whereas \SB9\ gives a short one.
\item Too low S/N: although the fit is improved with the orbital model, the astrometric signature of the secondary is below the noise level (cf. next section and Pourbaix (2001))
\end{itemize}
Jorissen et al. (this volume) present a complementary application to Ba stars.

\section{Why do so many systems get discarded?}

In the previous section 733 systems got discarded out of 863.  Although that is already a lot, lots of the kept ones are still troublesome.  How does the strength of these filters compare with each other (why does one discard HIP~112158) and what might me the missing filter (how to get rid of HIP~4584)?

Figure \ref{Fig:9periods} (right panel) indicates that 75\% of the systems are discarded because there is no improvement of the fit whether one adopts the orbital model or not.  The distribution of the orbital periods largely explains such a high percentage.  For instance, 363 systems have periods below 25 days thus making any astrometric wobble caused by the secondary unlikely to be detected by Hipparcos.  Systems with large orbital periods are also likely to suffer the same problem, the correlation among parameters (efficiency quite low) discarding some more.  Imposing $S/N>1.5$ mas would discard 35 additional systems (e.g. HIP~4584).  

Discarding systems like HIP~112158 was not a mistake per say but rather a way for the statistical tests to tell us something about that system.  Even though it is listed as an SB1 in \SB9, the discrepancy between the genuine Thiele-Innes and Campbell solutions reveals that Hipparcos did see the secondary.

\section{Conclusions}

The exploration of the richness of the Hipparcos observations is far from over.  Besides the new scientific results one can still derive from them, the readiness and free availability of these observations make them an excellent testbed for the preparations of some future astrometric missions.

\acknowledgments AJ and DP are Research Associates, FNRS (Belgium).  This research was supported in part by an ESA/PRO\-DEX Research Grants 90078 and 15152.

\end{document}